\documentstyle[twoside,fleqn,espcrc2]{article}

\newcommand{\be}{\begin{equation}}
\newcommand{\ee}{\end{equation}}
\newcommand{\ba}{\begin{eqnarray}}
\newcommand{\ea}{\end{eqnarray}}

\newcommand{\AmS}{{\protect\the\textfont2
  A\kern-.1667em\lower.5ex\hbox{M}\kern-.125emS}}

\title{Lifetimes of the $b$--flavored baryons in the light--front quark model
\thanks{The talk presented at the 4$^{th}$
International Conference on Hyperons, Charm and Beauty Hadrons,
Valencia, June 27--30, 2000. This work was supported in part by the
RFBR grants,
ref. No 00--02--16363 and 00-15--96786}}

\author{ I.M. Narodetskii
       \address{ITEP, 117218
       Moscow, Russia}}

\begin{document}

\begin{abstract}

\noindent The calculation of lifetimes of heavy--flavored baryons in the light--front quark model approach is briefly reviewed.

\end{abstract}

\maketitle

\section{INTRODUCTION}
\noindent The lifetimes of the $b$ flavored hadrons $H_b$ are related both to the CKM matrix
elements $|V_{cb}|$ and $|V_{ub}|$ and to dynamics of $H_b$ decays.
Weak decays of heavy hadrons involve three fundamental scales, the weak interaction scale $M_W$, the $b$--quark mass $m_b$, and the QCD scale $\Lambda_{QCD}$, which are strongly ordered: $M_W\gg m_b\gg \Lambda_{QCD}$. The first inequality means that the decay width of the $b$ quark rises so rapidly with $m_b$, namely
$\Gamma_b\sim G_F^2m_b^5$, that the {\it spectator process} dominates for large $m_b$.
The second inequality means that for large $m_b$ 
$b$--quark decays before it can  hadronize.
Thus in the limit $m_b\to \infty$ the lifetimes of all $b$ hadrons must
be equal: $\tau_{B^+}=\tau_{B_d^0}=\tau_{B_s^0}=\tau_{\Lambda_b}$.

Inclusive $H_b$ decays
can be treated with the help of an operator product expansion (OPE) combined with
the heavy quark expansion \cite{BSU97}. The OPE approach predicts that all corrections to
the leading QCD improved parton terms appear at the order $1/m_b^2$ and beyond.
Thus mesons and baryons containing b quark are expected to have lifetimes differing by no
more than a few per cent.
The result of this approach for the $\Lambda_b$ lifetime is puzzling because it
predicts that \cite{NS97}
$\left(\tau_{\Lambda_b}/\tau_B\right)_{OPE}=0.98+{\cal O}(1/m^3_b)$,
whereas the
experimental findings suggest a very much reduced fraction \cite{J97}
\be
\label{1}
\left(\tau_{\Lambda_b}/\tau_B\right)_{exp}=0.78 \pm 0.04,
\ee
or conversely a very much enhanced decay rate.

Application of OPE to decays of b--hadrons has potentially
two caveats. One is that OPE is used in Minkowsky  domain, and therefore
relies on assumption of quark--hadron duality at the energies involved in decays.
Another uncertainty arises from poor knowledge of matrix elements
of four--quark operators arising as terms of OPE. The decay rates of B and
$\Lambda_b$ are $\Gamma(B)=0.63\pm 0.02~{\rm ps}^{-1}$  and
$\Gamma(\Lambda_b)=0.83\pm~0.05~{\rm ps}^{-1}$ differing by $\Delta\Gamma(\Lambda_b)=
0.20\pm~0.05~{\rm ps}^{-1}$. The four--fermion processes of weak scattering and
Pauli interference calculated using the factorization approach and the description
of the baryon relying on quantum mechanics of only the constituent quarks
could  explain, under certain conditions, only $(13\pm 7)\%$ of $\Delta\Gamma(\Lambda_b)$.
\cite{R96}. However, the "spectator effects", which formally appear at ${\cal
O}(1/m_b^3)$ in OPE can be
considerably larger. A piloting lattice study of Ref. \cite{DSM99} suggests
that the effects of weak scattering and interference can eliminate
$\sim ~50\%$ of the discrepancy between the theoretical
prediction for $\tau_{\Lambda_b}/\tau_{B}$ and the experimental
result. It also appears that not all of the discrepancy can be accounted for by spectator effects. The OPE prediction, given all the uncertainties involved,
is unlikely to produce a number lower that $\sim ~0.9$.

It is not clear whether the present contradiction between
the theory and the data on $\tau_{\Lambda_b}/\tau_B$ is a temporary difficulty
or an evidence of fundamental flaws in theoretical understanding.
In spite of great efforts of experimental activity the $\Lambda_b$ lifetime remains
significantly low which continues to spur theoretical activity.
In this respect, the use of
phenomenological models, like the constituent quark model, could be of
interest as a complimentary approach to the OPE resummation method.
In this talk I shall consider the preasymptotic effects for the
$\Lambda_b$ lifetime in the framework of the
light--front (LF) quark model.
My talk is preceded by that of Dr.Kalman \cite{SK00}, covering, in part, the
same topic. Therefore, I will avoid those aspects that have been already discussed. 
Instead, I will concentrate on the calculation of lifetimes of $\Lambda_b$ and other $b$--flavored baryons.

\section{LF QUARK MODEL}
The LF quark model \cite{J99} is a relativistic constituent quark
model based on the LF formalism. In Ref. \cite{KNST99} this formalism has been
used  to establish a simple quantum mechanical relation between the inclusive  decay
rate of the B meson and that of a free ${\rm b}$ quark. The approach of \cite
{KNST99} relies on the idea of duality in summing over the final hadronic states.
It has been assumed that the sum over all possible charm final states $X_c$ can be
modeled by the decay width of an on--shell $b$ quark into on--shell $c$ quark
folded with the $b$--quark distribution function
$f_B^b(x,p^2_{\bot})=|\psi_B^b(x,p^2_{\bot})|^2$.
The latter represents the probability to find $b$ quark carrying
a LF fraction $x$ of the hadron momentum and a transverse relative momentum
squared $p^2_{\bot}$. For the partial rates $d\Gamma(B)/dq^2$ the above mentioned relation
takes the form
\be
    \label{2}
    {d\Gamma(B) \over dq^2} = {d\Gamma^b \over dq^2}R_B(q^2),
\ee
where $d\Gamma^b/dq^2$ is the free quark semileptonic or nonleptonic
differential decay rate, with
$q$ being the 4--momentum of the $W$ boson. The function
$R_B(q^2)$ incorporates the non perturbative
effects related to binding effect and to the Fermi motion of the heavy quark
inside the hadron
and is expressed in terms of 
the bound--state factor $|\psi_B^b(x,p^2_{\bot})|^2$. For the details see
\cite{KNST99}. The non--perturbative corrections to the free quark decay
rate vanish as $m_b\to\infty$ \cite{KNST99} but may be essential at finite values of the $b$
quark mass due to the difference between $m_b$ and $m_{H_b}$ and primordial motion of the $b$
quark inside $H_b$.

To define the LF wave function we use the simple operational
anz\"atz \cite{C92} and express
$\psi_B^b(x,p^2_{\bot})$
in terms of the equal time radial wave function
$\psi_B(p^2)$. Explicit formulae relating
$\psi_B^b(x,p^2_{\bot})$ and
$\psi_B(p^2)$ can be found {\it e.g.} in Ref. \cite{GNST97}.

\section{$B$--MESON LIFETIME}
In what follows the B meson orbital wave function
is assumed to be the Gaussian function as $
\psi_B(p^2)=\left(1/\beta_{b\bar d}\sqrt{\pi}\right)^{\frac{3}{2}}
\exp\left(-p^2/2\beta^2_{b\bar d}\right)$, where the parameter $1/\beta_{b\bar d}$
defines the confinement scale. We take  $\beta_{b\bar d}=0.45~{\rm GeV}$.
For $|\Psi_B(0)|^2$, the square of the
wave function at the origin,
we have $|\Psi_B(0)|^2=1.64\times 10^{-2}~{\rm GeV^3}$ that compares favorably
with the estimation in the constituent quark ans\"atz \cite {S82}
$|\Psi_B(0)|^2=M_Bf^2_B/12
=(1.6\pm 0.7)\times 10^{-2}~{\rm GeV}^3$
for $f_B=190~{\rm MeV}$.

We have found that the $\tau_{\Lambda_b}/\tau_B$ ratio is  rather stable
with respect to the precise values
of the heavy quark masses $m_b$ and $m_c$ provided $m_b-m_c\ge 3.5~{\rm GeV}$. From now on we
shall use the reference values $m_b=5.1~{\rm GeV}$ and $m_c=1.5~{\rm GeV}$.
The value of the CKM parameter
$|V_{cb}|$
cancels in the ratio $\tau_{\Lambda_b}/\tau_{B}$, but is important
for the absolute rates. Details of our calculations of $\Gamma(B)$
are given in Table 1 for the three values of the
constituent mass $m_{sp}$. The values $m_{sp}\sim 300~(200)~{\rm MeV}$ are usually
used in non-relativistic (relativistic) quark models.
We have also
considered a very low constituent quark mass $m_{sp}=100~{\rm MeV}$ to see how
much we can push up the theoretical prediction for  $\Gamma(\Lambda_b)
/\Gamma(B)$, see below. All the semileptonic widths include
the pQCD correction as an overall reduction factor equal to 0.9.
Following Ref. \cite{GNST97} the transitions to baryon-antibaryon
$(\Lambda_c\bar N~{\rm and}~\Xi_{cs}\bar \Lambda)$ pairs are included.
In addition we have added ${\rm BR}\approx 1.5\%$ for the
$b\to u$ decays with $|V_{ub}/V_{cb}|\sim 0.1$.
The value of
$|V_{cb}|$ is defined by the condition that the calculated B lifetime
is $1.56~ps$.

\begin{table}[t]
\caption{
The branching fractions (in per cent) for the inclusive B decays 
calculated within the LF quark model for the
several values of $m_{sp}$. The values of $|V_{cb}|$ in units of
$10^{-3}\sqrt{1.56~{\rm ps}/\tau^{(exp)}(B)}$ are also reported.}
\begin{center}
\begin{tabular*}{70mm}{@{}l@{\extracolsep{\fill}}rrrrr}
\hline $m_{sp}[{\rm MeV}]$ & $100$ & $200 $ & $300$ \\ \hline\hline $b\to ce\nu_e $
& 10.65 & 10.98 & 11.46 \\ $b\to c\mu\nu_{\mu} $ & 10.59 & 10.93 &
11.40 \\ $b\to c\tau\nu_{\tau} $ & ~2.47 & ~2.51 & ~2.57 \\ \hline
$b\to cd\bar u $ & 47.88 & 47.88 & 47.52 \\ $b\to c{\bar c}s $ &
14.07 & 14.31 & 14.63 \\
$b\to cs\bar u$ & ~2.94 & ~3.09 & ~3.32 \\ \hline
$B\to \Xi_{cs}\bar \Lambda_c $ & ~2.22 & ~1.83
& ~1.43 \\ $B\to \Lambda_c \bar N $ & ~7.70 & ~6.91 & ~6.02 \\ \hline
$b\to u $ & ~1.47 & ~1.54 &
~1.66 \\ \hline\hline $|V_{bc}|$ & 38.3 & 39.3 & 40.7 \\ \hline
\end{tabular*}
\end{center}
\end{table}

\section{$\Lambda_b$ LIFETIME}
We shall analyze the inclusive semileptonic and non--leptonic  $\Lambda_b$
rates on simplifying assumption that $\Lambda_b$ is composed of a heavy quark
and a light scalar diquark with the effective mass $m_{ud}$.
Then the treatment of the inclusive $\Lambda_b$ decays is simplified to a
great extent and one can apply the model considered above with the minor modifications.
For the heavy--light diquark wave function $\psi_{\Lambda_b}$
we again assume the Gaussian ans\"atz  with the
oscillator parameter $\beta_{bu}$. The width of $\Lambda_b$ can be obtained from that
of $B$ by the replacements
$M_B\to M_{\Lambda_b}$, $m_{sp}\to m_{ud}$ and $\beta_{b\bar d}\to \beta_{bu}$.
The latter two replacements change $f_{\Lambda_b}^b$,
the $b$ quark distribution function inside the $\Lambda_b$, in comparison with $f_B^b$.

The inclusive nonleptonic channels for $\Lambda_b$ are the same as for B meson
except the decays into baryon--antibaryon pairs which are missing in case
of $\Lambda_b$. The absence of this decay channel leads to the reduction of
$\Gamma(\Lambda_b)$ by $\approx 7\%$.
This effects prolongs $\tau_{\Lambda_b}$ over $\tau_B$. The phase space enhancement
in $\Lambda_b$ is marginal and can not be responsible for a shorter lifetime of
$\Lambda_b$. The only distinction
between the two lifetimes, $\tau_{\Lambda_b}$ and $\tau_B$,
can occur  due to the difference of the binding and Fermi motion effects. These effects are
encoded in the parameter $x_0=m_b/m_{H_b}$ and the distributions $f_{\Lambda_b}^b$
and $f_B^b$.


The concrete amount of the deviation between $\tau_{\Lambda_B}$ and $\tau_B$
depends in essential way on 
$m_{ud}$ and $\beta_{ub}$. In what follows the diquark mass is taken as
\be
\label{dm}
m_{ud}=m_*=\frac{1}{2}(m_u+m_d-m_{\pi}).
\ee
In this relation
inspired by the quark model,
the factor $1/2$ arises from the different color factors for $u$ and $\bar d$
in the $\pi$--meson  ( a triplet and antitriplet making a singlet) and
$u$ and $d$ in the the $\Lambda_b$ (two triplets making an antitriplet).

The quantity $\beta_{ub}$ can be translated into the ratio
of the squares of the wave functions determining the probability to find a light
quark at the location of the $b$ quark inside the $\Lambda_b$ baryon and $B$
meson, {\it i.e.}
$r=|\Psi_{\Lambda_b}(0)|^2/|\Psi_B(0)|^2=
\left(\beta_{bu}/\beta_{b\bar d}\right)^3$.
If we take $r=1.2\pm~0.2$ as suggested by the lattice study of Ref. \cite{DSM99}
we obtain
$\left(\tau_{\Lambda_b}/\tau_B\right)_{LF}=0.88\pm~0.02$ for $m_{sp}=200~{\rm MeV}$.
Adding the
contribution of the 4--quark operators $\Delta\Gamma^{4q}=0.08\pm~0.02$
from \cite{DSM99} we reproduce the experimental result (\ref{1}).


It is instructive to note that the calculated branching fractions of $\Lambda_b$  show marginal
dependence on the choice of the model parameters; they are $\sim 11.5~\%$
for the semileptonic $b\to ce\nu_e$ transitions,
$\sim 2.8\%$ for  $b\to c\tau\nu_{\tau}$, $\sim 50\%$ for the nonleptonic
$b\to cd\bar u$ transitions, and $\sim 16\%$ for $b\to c\bar cs$
transitions.

An exploratory study of the
dependence of $\tau_{\Lambda_b}$ on $m_{ud}$ and $r$ has been performed in
Ref. \cite{KNO00}. Two choices for diquark masses were taken:
$m_{ud}=m_u+m_d$ corresponding to zero binding approximation and $m_{ud}=m_*$
from (\ref{dm}). The wave
function ratio has been varied in the range $0.3\le r\le 2.3$
\footnote{
Recall that estimates of the parameter $r$ using the non relativistic quark model or
the bag model  
or QCD sum rules
are typically in the range $0.1-0.5$. The value $r=2.3$ corresponds to the
upper Rosner estimation \cite{R96} obtained from the ratio of hyperfine splitting between $\Sigma_b$ and
 $\Sigma_b^*$ baryons and $B$ and $B^*$ mesons .} that
corresponds to
$0.3~{\rm GeV}\le \beta_{bu}\le 0.6~{\rm GeV}$. The results
for $\tau_{\Lambda_b}/\tau_{B}$
are exhibited in
Table 2, which we now discuss.

\begin{table}[t]
\caption{The LF quark model results for the ratio
$\tau_{\Lambda_b}/\tau_{B}$ calculated for
different values of $r$ and $m_{sp}$. $m_{sp}$ are in units of MeV. 
$m_{ud}$ are given by (\ref{dm}). 
The results corresponding to $m_{ud}=m_u+m_d$ are given in parentheses.
}

\begin{center}
\begin{tabular*}{70mm}{@{}l@{\extracolsep{\fill}}rrrrr}

\hline
 $r
\setminus m_{sp}$ & 100 & 200 & 300\\ \hline\hline
0.3 & 0.80 (0.81)& 0.79 (0.84) & 0.80 (0.88)
 \\ \hline
1.0 & 0.87 (0.88)& 0.86 (0.90)& 0.86 (0.93)\\ \hline
2.3 & 0.93 (0.94)& 0.91 (0.95) & 0.91 (0.97)

 \\ \hline\hline

\end{tabular*}
\end{center}
\end{table}

We observe that to decrease the theoretical prediction for
$\tau_{\Lambda_b}$ requires to decrease the value of the hadronic parameter
$r$  to 0.3-0.5 and the value of $m_{sp}$ to $\sim 100~{\rm MeV}$.
For example, assuming that $r\sim 0.3$ we find
that  the lifetime ratio is decreased from $0.88$ to $0.81$ if
$m_{sp}$ is reduced from $300~{\rm MeV}$ to $100~{\rm MeV}$ and $m_{ud}=m_u+m_d$. 
For the 
$m_{ud}= m_*$
the ratio is almost stable ($\sim
0.8$), so that reducing of the diquark mass produces a decrease
of the lifetime ratio by $1\%$, $5\%$, and $8\%$ for $m_{sp}=100,~200,~
{\rm and}~300~{\rm MeV}$, respectively.
Varying the spectator quark mass in a similar way we find that for
the "central value" $r\sim 1$ the lifetime ratios
are reduced from $0.93$ to 0.88 for $m_{ud}=m_u+m_d$ and remain almost
stable ($\sim 0.86$) for $m_{ud}\sim m_*$.
For the largest possible value of $r$ suggested in \cite{R96},
$r\sim 2.3$, the lifetime ratios
are reduced from 0.97 to 0.94 in the
former case and remain almost stable $\sim 0.91-0.93$ in the latter
case.

The lifetimes of other $b$--flavored baryons are reported in Table 3.
The results are
obtained using $r=1$ and adding the marginal contribution of $\Delta\Gamma^{4q}$ from Ref. \cite{R96}.
As we have  discussed earlier this contribution may be strongly enhanced by the QCD
effects in which
case the lifetime predictions should be decreased by $\sim 10\%$.

\section{CONCLUSIONS}
If the current value of
$\left(\tau(\Lambda_b)\right)_{exp}$ persists, the most likely 
explanation is that some hadronic matrix elements of four--quark operators
are larger than the naive expectation \cite{NS97}.
If a significant fraction $\sim50\%$ of the discrepancy between the theoretical
prediction for $\tau_{\Lambda_b}/\tau_{B}$ and the experimental
result can be accounted for the spectator effects \cite{DSM99} then
the remainder of the discrepancy can be naturally explained by the preasymptotic
effects.
We
have found that the binding effect produces for
$r= 1.2\pm 0.2$ $m_{sp}=200~{\rm MeV}$ and $m_{ud}=250~{\rm MeV}$ an additional
increase of $\Delta\Gamma(\Lambda_b)/\Gamma(B)$ by $12\pm 2\%$
\footnote{Note that for these values of quark and diquark
masses the Fermi motion effects in $\Lambda_b$ and $B$ do not differ significantly.}
.

\section*{ACKNOWLEDGMENTS}

\noindent I would like to thank Calvin Kalman, Pepe Salt  and Miguel--Angel Sanchis--Lozano
for organizing an excellent Conference with a stimulating scientific
program.


\begin{table}[t]
\caption{The predicted lifetimes $\tau_{H_b}$ (in units  of ps). $r=1$. The constituent quark
masses are $m_b=5.1~{\rm GeV}$, $m_c=1.5~{\rm GeV}$, $m_u=0.2~{\rm GeV}$, $m_s=0.35~{\rm GeV}$.
The diquark masses are $m_{ud}=0.25~{\rm GeV}$, $m_{us}=0.55~{\rm GeV}$, $m_{ss}=0.7~{\rm GeV}$.
$|V_{cb}|=0.0393$.
}

\begin{center}
\begin{tabular*}{70mm}{@{}l@{\extracolsep{\fill}}rrrrr}

\hline
$H_Q$ & $~~\tau(H_Q)$ &  $\tau_{exp}(H_Q)$ \\
\hline\hline
$B^0$ & $1.56$ & $1.56\pm 0.04$ \\
\hline
$B_s$ & $1.53$ & $1.54\pm 0.07$\\
\hline\hline
$\Lambda_b$ & $1.31$ & $1.24\pm 0.08 $\\
\hline
$\Xi_b^0$ & $1.31$ & $1.39^{+0.34}_{-0.28}$\\
\hline
$\Xi_b^-$ & $1.41$ & \\
\hline
$\Omega_b$ & $1.32$ & \\

\hline
\end{tabular*}
\end{center}
\end{table}


\end{document}